\documentstyle[epsfig]{jaa}

\begin{document}

\pubyear{2001}
\volume{22}
\pagerange{1--8}
\date{Received 2000 October 27; accepted 2001 February 05}

\title [GMRT detection of HI absorption in A2125 galaxy]
       {GMRT detection of HI 21 cm-line absorption \\
       from the peculiar galaxy in Abell 2125} 

\author[Dwarakanath \& Owen]
       {K. S. Dwarakanath$^1$\thanks{e-mail:dwaraka@rri.res.in} and F. N. Owen$^2$
        \thanks{e-mail:fowen@nrao.edu} \\
       $^1$Raman Research Institute, Bangalore 560 080, India \\
       $^2$National Radio Astronomy Observatory, Socorro, NM 87801, USA}

\maketitle

\begin{abstract}
Using the recently completed Giant Meterwave Radio Telescope, 
we have detected the HI 21 cm-line absorption from the peculiar galaxy C153 in the
galaxy cluster Abell 2125. 
The HI absorption is at a redshift of 0.2533, with a peak optical depth of 0.36. The full width
at half minimum of the absorption line is 100 km s$^{-1}$.  The estimated column density of
atomic Hydrogen is 0.7$\times$10$^{22}$(T$_{s}$/100) cm$^{-2}$. The HI absorption
is redshifted by $\sim$ 400 km s$^{-1}$ compared to the [OIII] emission line from this system. 
We attribute this to an in-falling cold gas, or to an out-flowing ionised gas, or to a combination
of both as a consequence of tidal interactions of C153 with either a cluster galaxy or the cluster
potential.
\end{abstract}


\section{Introduction}

Abell 2125 is a rich cluster of galaxies (Abell Class 4) at a redshift of 0.246. 
Recently a detailed optical and radio study of this cluster was 
published (Dwarakanath \& Owen 1999, Owen et al 1999). These studies have detected 27 radio 
galaxies in this cluster above a 20 cm luminosity limit of 1.4 $\times$ 10$^{22}$ W Hz$^{-1}$
(H$_{o}$ = 75 km s$^{-1}$ Mpc$^{-1}$, q$_{o}$ = 0.1). In projection, these radio galaxies extend 
over $\sim$ 5 Mpc along a band running from the northeast to the southwest of the cluster center.
About half of these galaxies show signs of star formation, with the largest concentration of them
in the southwest clump $\sim$ 2 Mpc in projection from the cluster center. There is a bimodal distribution
of the 27 cluster members in radio luminosity, with the majority below a spectral luminosity at 20 cm
of 10$^{23}$ W Hz$^{-1}$. Star formation is primarily responsible for the radio emission of these
galaxies. Rest of the members with a spectral luminosity above this value owe most of their radio emission
to an AGN activity in them. 

In Fig. 1 a radio image of the central region of the cluster is reproduced 
from Dwarakanath \& Owen (1999).
All the sources in this figure are cluster members. The brightest source in this figure (C153) has a peak
flux density of 23.2 mJy/beam and is within 30$''$ ($\sim$ 100 kpc) in projection from  the cluster core.
This is the second brightest radio source in the cluster with a 20 cm spectral luminosity
of 3.3 $\times$ 10$^{24}$ W Hz$^{-1}$. Most of this radio emission arises in the core of
this galaxy. The core is essentially unresolved at a resolution of 
$\sim$ 1.8$''$. However, faint extended radio emission can be seen towards 
northwest and southeast directions. This galaxy has a ratio of [NII] to H$_{\alpha}$ consistent
with an AGN (Owen et al 1999). But, the value of the continuum break around 
4000 \AA (D(4000)), and 
the value of B--R (color) are very small and consistent with star formation
as well as AGN activity. The star formation rates implied by the H$_{\alpha}$ and O[II] lines 
are consistent with each other but are a factor of 50 too small to account for the observed
radio emission (Owen et al 1999). 
This discrepancy can arise either because the star formation rates are 
under-estimated due to dust in C153, or most of the radio emission from C153 is due to an AGN
activity. In comparison with the blue luminosities and colors of other cluster members where
the star formation rates estimated by the optical lines and the radio continuum agree, it
is apparent that dust obscuration is unlikely to be significant in C153. 
Most of the radio emission from C153 must therefore be due to an AGN activity at its center.

\begin{figure*}
\epsfig{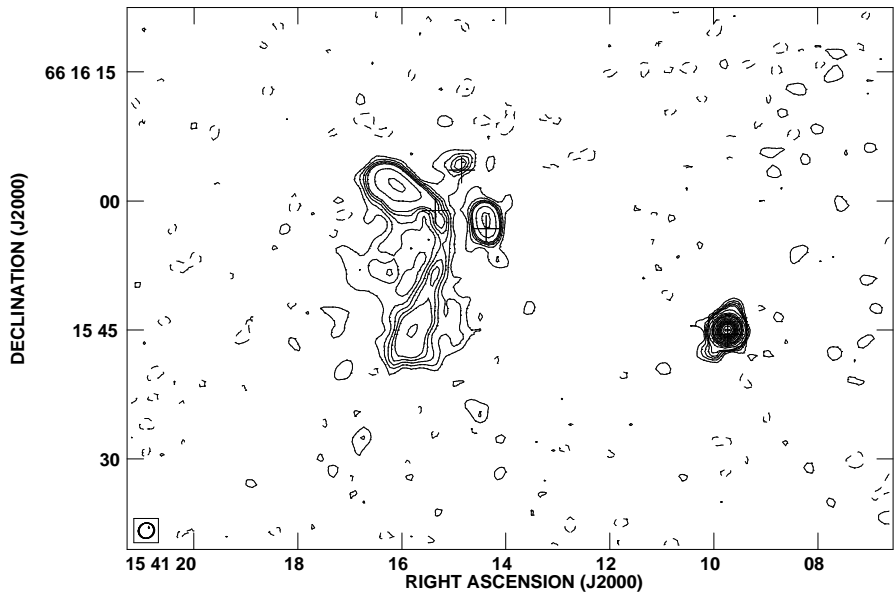}
\caption{ Radio image of the center of Abell 2125. This is a 20 cm image made with the VLA
in its A configuration with a resolution $\sim$ 1.8$^{''}$. The crosses mark the positions 
of optical identifications. The brightest source in this figure (C153) has a peak continuum
surface brightness  
of 23.2 mJy/beam and is within $\sim$ 100 kpc (in projection) from the cluster core. 
This source shows signs of star formation although it is an AGN. The rms
in this image is $\sim$ 30 $\mu$Jy/beam. The contours are in steps of 60 $\mu$Jy/beam from
-120 to 300 $\mu$Jy/beam (excluding 0), in steps of 0.6 mJy/beam from 0.6 to 3 mJy/beam, 
in steps of 1.5 mJy/beam from 4.5 to 15 mJy/beam, and in steps of 3 mJy/beam from 18 to 
30 mJy/beam. The synthesised beam is at the bottom left corner.} 
\end{figure*}

Since there is star formation and AGN activity in C153, we might expect to detect some of the
cold gas feeding the AGN. Such a cold gas could be in a torus close to the nucleus. Alternatively,
we might expect to detect the HI in the interstellar medium of the host galaxy. 
Such a system with multiple
HI absorption components was indeed detected in Hydra A (Dwarakanath, Owen, \& van Gorkom 1995). 
In-fall velocities of HI absorption components up to 400 km s$^{-1}$ have been detected in
radio elliptical galaxies (van Gorkom et al 1989, van der Hulst et al 1983, Shostak et al 1983). 
This in-falling HI gas has been attributed to cold gas loosing its angular momentum due to
frictional drag on it by the pressure supported hot gas in the ellipticals. 
Interactions and mergers are quite common in cluster galaxies (Lavery \& Henry 1994, Dressler et al
1994, Couch et al 1994, Wirth, Koo, \& Kron 1994). Such interactions can lead to disturbed morphologies
and chaotic velocity fields of the interacting galaxies. In-flow and/or out-flow of ionised 
and/or atomic gas is expected from such galaxies. It is interesting to explore if C153 displays
any such phenomena.  

\section{GMRT Observations}

The Giant Meterwave Radio Telescope (GMRT) located at Khodad near Pune in India, is an aperture  
synthesis instrument consisting of 30 
fully steerable parabolic dishes of 45 m diameter. Six antennas are distributed along
each of the three arms of a rough Y and the remaining 12 antennas are more or less randomly
placed in a compact cluster near the center of the Y. The center of the Y, called the 
$'$Central Square$'$ has a minimum baseline of $\sim$ 100 m and a maximum baseline of $\sim$ 1 km. 
The maximum baseline of the GMRT array is $\sim$ 25 km (Swarup et al 1991). 
Currently the antennas have dual 
polarised feeds at 5 frequencies viz., 150, 230, 327, 610, and 1420 MHz. 
The 1420 MHz feed is a broad-band corrugated horn covering the frequency range of 900 - 1450 MHz
continuously. The full width at half maximum of the primary beams of the 
GMRT dishes at 20 cm is $\sim$ 25$'$. 

For the current observations the antennas were pointed at $\alpha$(J2000) = 15$^{h}$41$^{m}$09$^{s}$
and $\delta$(J2000) = 66$^{o}$15$'$44$''$.
The backend used was a 30 station FX correlator which produces 128 spectral channels across the chosen
baseband bandwidth. Any bandwidth which is 2$^{n} \times$ 62.5 KHz with n taking integer
values between 0 and 8 can be chosen. 
In the present observations, a bandwidth of 8 MHz centered at 1134 MHz was used. 
This 8 MHz band covered a velocity
range of $\sim$ 2000 km s$^{-1}$ at the redshift of 0.2526 of C153. The corresponding velocity
resolution is $\sim$ 15.7 km s$^{-1}$. Although most of the 30 antennas were included in the 
observations, the total number of antennas that were used in the final analysis was 14 with
a maximum baseline of $\sim$ 2.5 km. The
integration time on the source was $\sim$ 4 Hr. Gain and bandpass calibrations were carried out 
by observing the VLA calibrator 1634+627 once an hour. 
The absolute flux densities were estimated by observing 1331+305 (3C 286). 

The raw data from the telescope was converted to FITS format and analysed using the Astronomical Image
Processing System (AIPS -- developed by the National Radio Astronomy Observatory). 
The channels from the flat portion of the spectrum and which were free from line absorption 
were used to estimate
the continuum visibilities. Spectral-line visibilities in each of the spectral channels 
were obtained by subtracting the estimated continuum contribution from the observed visibilities. A continuum
image and a spectral cube were made using AIPS. The synthesised beam was 36$''$ $\times$ 22$''$ at a position
angle of -22$^{o}$.  The r.m.s. value in the spectral cube was 1.5 mJy/beam/channel, close to the
 expected value. 
The spectral cube was featureless except for the absorption line in the galaxy C153. The spectral
cube was Hanning smoothed with a resulting r.m.s. of 0.9 mJy/beam/channel. The spectrum towards
C153 from this cube is displayed in Fig. 2. The velocity resolution in this spectrum is $\sim$ 32 km s$^{-1}$.
 
The HI absorption line is at a redshift of 0.2533, or, at a systemic velocity of 66550 km s$^{-1}$.
The best-fit Gaussian gives a full width at half minimum of 100 km s$^{-1}$. The peak optical
depth is 0.36 and the corresponding HI column density is 0.7 $\times$ 10$^{22}$(T$_{s}$/100) cm$^{-2}$.
The corresponding HI mass in a sphere of radius 1 kpc is $\sim$ 10$^{8}$M$_{\odot}$.

\begin{figure*}
\epsfig{file=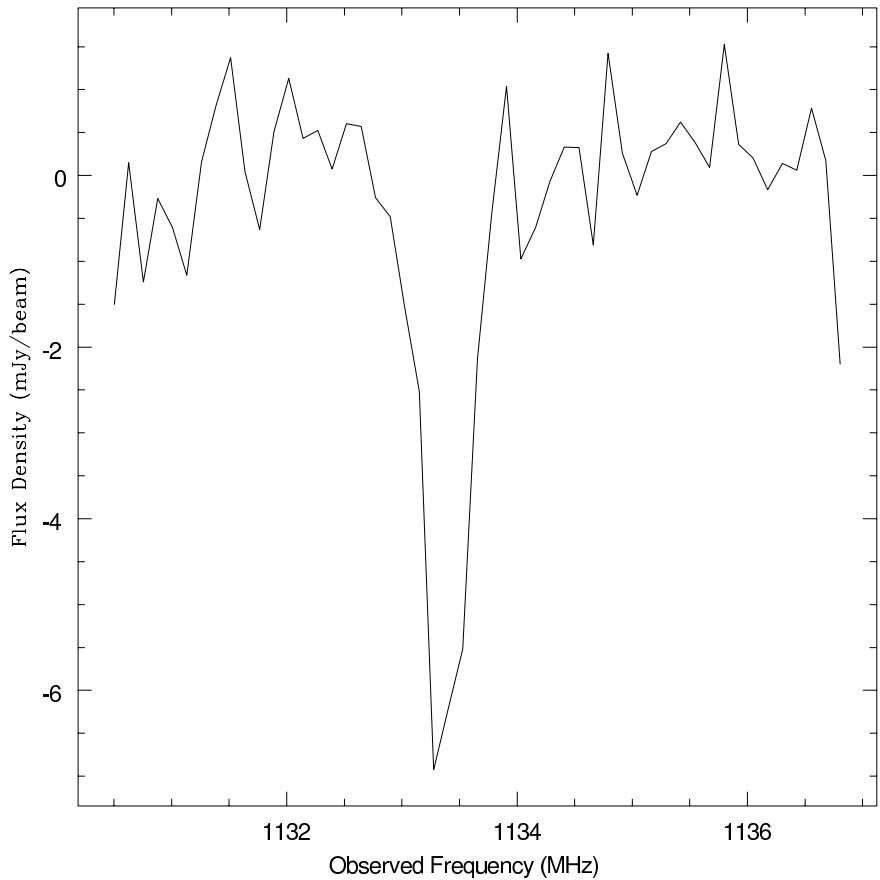, width= 12cm}
\caption{ The HI 21 cm-line absorption in C153 obtained using the GMRT. The zero on the y-axis corresponds
to the
peak continuum surface brightness of C153 of 24.1$\pm$1.5 mJy/beam and is in excellent agreement with
23.2$\pm$0.075 mJy/beam estimated from earlier VLA observations 
(Dwarakanath \& Owen 1999). The spectrum was obtained at
$\alpha$(J2000) = 15$^{h}$41$^{m}$09.83$^{s}$, and $\delta$(J2000) = 66$^{o}$15$^{'}$44.1$^{''}$.
The original spectrum was Hanning smoothed and alternate channels plotted to obtain the spectrum
shown here with a velocity resolution of $\sim$ 32 km s$^{-1}$. The peak absorption has an optical
depth of 0.36 and corresponds to
a redshift of 0.2533. The [OIII] emission line from this system is blue shifted w.r.t. the HI absorption
line by $\sim$ 400 km s$^{-1}$.} 
\end{figure*}

\section{Discussion}

The most interesting aspect of the HI absorption in C153 is that it is redshifted by
$\sim$400 km s$^{-1}$ w.r.t. the velocity of the [OIII] emission line, 
but is within $\sim$60 km s$^{-1}$
of the H$_{\alpha}$  and [OII] emission lines (Owen, under preparation). 
The errors in the redshift estimates based
on the optical lines are expected to be $\sim$60 km s$^{-1}$ making the velocity difference between the HI
absorption line and the [OIII] emission line significant. The [OII] and the [OIII]  line emission
appear to come from a more extended, disturbed region than the HI absorption and the H$\alpha$
emission lines.  There can be different scenarios accounting for the observed differences in
the velocities and in the spatial extents of different lines. 
There can be an outflow of [OIII] emitting gas while the HI absorption, the H$\alpha$, and
the [OII] lines are at rest w.r.t. the galaxy,  
or there can be an in-fall of the gas seen in HI absorption, H$\alpha$, and [OII] lines while 
the [OIII] emitting gas is at rest,  or there can be a combination of these two situations. 
Any model that attempts to explain the current observations should also account
for the width, and the HI column density of the absorption line. We consider
the following physical models and discuss their plausibilities in the context of the 
current observations :  (a) a gas torus close to the AGN, (b) the interstellar
medium of the host galaxy, (c) an interacting, intervening cluster galaxy, and 
(d) tidal interactions with the cluster potential.

If the HI absorption is due to a gas torus surrounding the AGN, both the width, and the
column density of the HI line can be explained. A rotating gas torus of dimensions comparable 
to the linear size of the central radio continuum source can easily account for the observed width
of the HI absorption line. While the linear size of the radio continuum source is unknown, it is
expected to be less than 1 kpc based on the earlier observations with the VLA (Dwarakanath,
\& Owen 1999).  For e.g., a gas torus of HI density $\sim$ 20 cm$^{-3}$,  
and of extent $\sim$ 100 pc with a circular velocity
$\sim$ 250 km s$^{-1}$ surrounding a radio continuum source of comparable extent can account for
the observed width and column density of the HI absorption line. The HI absorption will 
be essentially at rest w.r.t. the AGN since there is no systematic motion involved
other than the circular motion of the torus. 
The H$\alpha$ emission could come from ionised part of 
the torus. However, the spatially extended [OII] and [OIII] emission must have a different
origin. Thus, a gas torus has a limited scope in explaining the current observations.  

If the HI absorption is due to the interstellar medium of the host galaxy, then, 
for an assumed spin temperature of 100 K, the total column density of HI in the line is comparable 
to the integrated HI column density along the line of sight at low galactic latitudes 
($\mid$b$\mid$ $<$ 5$^{o}$)
in the Galaxy (Dickey \& Lockman 1990). This similarity in the column densities indicates that 
the HI absorption in C153 could also be due to the line of sight gas in its galaxy.  
However, the observed HI column density requires a near edge-on alignment of the disk of the 
galaxy to our line 
of sight which has a rather small probability. In addition, since the circular
motion of the disk-gas is not expected to have any significant component of velocity
along the line of sight, the HI absorption is expected at a velocity not too far from
the systemic velocity.
It is possible to account for some inward velocities by gravitational in-fall of the cold HI 
gas due to loss of 
angular momentum. This loss of angular momentum is due to the frictional drag on the cold gas 
by the surrounding hot gas (Gunn 1979). This model has been adapted by van Gorkom
et al (1989) to account for the HI in-fall velocities observed in radio ellipticals.
In their adaptation of this model, they assume spherical cold clouds moving in circular orbits through
pressure supported hot gas. These clouds experience a frictional drag and acquire radial 
in-fall velocities. Geometrical
consistency demands that the radial velocity be not more than 1/(2$\pi$) times the
circular velocity. While small HI in-fall velocities $\sim$ 50 km s$^{-1}$ observed in some
radio ellipticals can be explained by this model, larger HI in-fall velocities
observed by them in these ellipticals must have a different origin. 
Note that there is one example (viz., NGC 315) of an HI absorption redshifted by $\sim$ 400 km s$^{-1}$ 
w.r.t. the systemic velocity (Shostak et al 1983). In this case, 
the column density is 4.6$\times$10$^{20}$ cm$^{-2}$ and the line width is $<$ 5 km s$^{-1}$.
These values are typical of interstellar clouds in the Galaxy. In the case of NGC315
one is witnessing a single cloud along the line of sight which is presumably in
a state of free-fall to its center. However, in the present case, the HI column density
is 15 times larger, and the line is 20 times wider. This cold gas presumably has a different origin. 
Even if we assume that the HI absorption is at rest w.r.t. the galaxy (C153), thus alleviating
the difficulties in accounting for its large in-fall velocities, the requirement of
near edge-on alignment of the galaxy to our line of sight to account for the observed HI
column density makes this model less probable. 

Absorption due to an intervening cluster galaxy alleviates the difficulties encountered
in the two scenarios mentioned earlier. Galaxies in clusters are known to undergo interactions,
and mergers. If a dwarf galaxy of HI mass $\sim$ 10$^{8}$ M$_{\odot}$, and of size a few kpc
 is interacting, and falling into C153, it can explain the current observations. 
The in-fall velocity of the dwarf will be at least as much as the escape velocity from C153.
The escape velocities from galaxies can be greater, or, of $\sim$ 400 km s$^{-1}$.
Such an in-falling dwarf can easily explain the observed redshift of the HI absorption
line, if we attribute the observed velocity difference between the HI absorption and 
[OIII] emission to an in-falling cold HI gas. The width of the HI absorption line simply results from the 
rotation of the dwarf. Given a sufficient size for this dwarf of a few kpc, its alignment
with the AGN  need not be as critical as in the earlier scenario (case (b)) to give rise to  
the HI absorption. A consequence of such an interaction is a disturbed morphology of the
galaxies. Indeed, C153 shows a disturbed morphology as seen in the HST images. In addition,
such an interaction can pull off gas from the outer regions of galaxies due to tidal
interactions, and cause an outflow of gas. This can also account for the observed difference
in the velocities of HI absorption and [OIII] emission lines. 
However, there is one drawback in this picture. 
The galaxy C153 is within $\sim$ 100 kpc in projection from the cluster core. If C153 is
indeed close to the cluster core, i.e., no projection effects, then a dwarf galaxy 
in-fall into C153 is less
likely in the core of a high velocity dispersion ($\sim$ 900 km s$^{-1}$) 
cluster like Abell 2125. 
This drawback is less serious if C153 were to be in the outer regions of the
cluster.

If C153 is not in the outskirts of the cluster but is indeed close to the core then we can 
advance the following alternative scenario. The observed  galaxy density distribution in Abell 2125 and  
its X-ray intensity distribution obtained from ROSAT PSPC  observations 
are very similar (Owen et al 1999). The galaxy C153 is within $\sim$ 100 kpc from the
peaks of these two distributions. Since C153 has a velocity $\sim$ 1500 km s$^{-1}$ w.r.t.
the cluster mean and is in the dense core of the cluster, both tidal stripping
and ram-pressure stripping are expected to be very effective. 
Recent simulations show that stripped gas quite often falls back onto the galaxy
(Barnes, \& Hernquist 1998, and references therein). 
In such a situation, redshifted HI absorption from the in-falling cold gas is
expected. Alternatively, the stripping can lead to an outflow of gas which is seen 
in the lines of [OIII] emission. Thus, this model can account for both in-falling cold
gas and out-flowing ionised gas. A consequence of tidal stripping and ram-pressure
stripping is a disturbed morphology of C153 which is seen in its HST images. Future
VLBA images in radio continuum and in the 21 cm-line of HI as well as imaging in 
the optical emission lines could throw more light on the nature
of this source.

%\acknowledgements
\section*{Acknowledgements}
The Giant Meterwave Radio Telescope is the result of dedicated efforts
of a large number of people at the National Center for Radio Astrophysics, Pune
and at the GMRT site, Khodad. N.V.G. Sarma and his team at the  
 Raman Research Institute, Bangalore were responsible for designing and building
the broad-band 21 cm feeds and receivers for the GMRT. We thank 
Rajaram Nityananda and K. R. Anantharamaiah for stimulating discussions and 
useful comments on the paper. 

%\clearpage

%\clearpage

\end{document}